\documentclass[b5paper,10pt]{article}
\usepackage{graphicx}
\usepackage{amsmath,amsfonts}

\usepackage{amssymb}
\usepackage{graphicx}
\usepackage{xspace}

\usepackage[utf8]{inputenc}
\usepackage[T1]{fontenc}
\usepackage{lmodern}

\begin{document}
\begin{center}
  {\Large\bf
    Stability and Bifurcation Analysis of Coupled Fitzhugh-Nagumo Oscillators
  }
\mbox{}\\[11pt]
{\large\sc
  William Hanan$^{1}$, Dhagash Mehta$^{1}$, Guillaume Moroz$^{2*}$, Sepanda Pouryahya$^{1}$
}
\mbox{}\\[5pt]
$^1$ Mathematical Physics Department, National University of Ireland Maynooth, Maynooth, Co. Kildare, Ireland \\
{\tt wgh@thphys.nuim.ie, dmehta@thphys.nuim.ie, sepour@thphys.nuim.ie}\\[5pt]

$^2$ Institut de Recherche en Communications et Cybern\'{e}tique de Nantes UMR CNRS 6597,  \\
1, rue de la Noe, BP 92101, 44321 Nantes Cedex 03 France \\
{\tt Guillaume.Moroz@irccyn.ec-nantes.fr}\\[5pt]

...........

\footnotetext[2]{Correspondence to: IRCCyN, 1, rue de la Noe, BP 92101, 44321 Nantes Cedex 03 France.  Tel : +33 2 40 37 69 00. Fax : +33 2 40 37 69 30 }

\end{center}

\begin{abstract}
Neurons are the central biological objects in understanding how the brain works. The famous Hodgkin-Huxley model, which describes
how action potentials of a neuron are initiated and propagated, consists of four coupled nonlinear differential equations.
Because these equations are difficult to deal with, there also exist several simplified models, of which many exhibit polynomial-like non-linearity. Examples of such models are the Fitzhugh-Nagumo (FHN) model, the Hindmarsh-Rose (HR) model, the Morris-Lecar (ML) model and the Izhikevich model. In this work, we first prescribe the biologically relevant parameter ranges for the FHN model and subsequently study the dynamical behaviour of coupled neurons on small networks of two or three nodes. To do this, we use a computational real algebraic geometry method called the Discriminant Variety (DV) method to perform the stability and bifurcation analysis of these small networks.  A time series analysis of the FHN model can be found elsewhere in related work \cite{Hanan10}.
\end{abstract}

\paragraph{The Fitzhugh-Nagumo Model}
The Fitzhugh-Nagumo (FHN) system of equations is a prototypic model of excitability. Introduced by FitzHugh \cite{fitzhugh61} with equivalent circuit by Nagumo et al.\cite{nagumo62} the system is a generalization of the Van der Pol oscillator \cite{vanderpol20} and a reduction of the neural electrophysiological model due to Hodgkin and Huxley (HH) \cite{hodgkinhuxley52}. As a generic model of excitability and oscillatory dynamical behavior, the FHN system is of relevance for a range of physical and physiological research topics
\cite{marino06,shoji05,marts07,lin06,barland03,descalzi09}, the most extensive of which are those concerned with cardiac 
\cite{maree97,biktashev02,sinha02} and neuronal 
\cite{brown99,toral03,wei08}
cell dynamics. The corresponding equations for $n$ electrically coupled FHN neurons are
\begin{equation}
\frac{d x_{i}}{d t} = x_{i} - \frac{x_{i}^{3}}{3} - y_{i} + g \sum_{j=1}^{n} (x_{i} - x_{j}), \hspace{0.5cm}
\frac{d y_{i}}{d t} = \epsilon (x_{i} + a - b\, y_{i})\label{eq:n_FHN_general_eq}
\end{equation}
for $i = 1, \dots, n$ and where we have taken $a \in [-2, 2]$, $b \in (0, \infty)$, $-1 \leq g \leq 1$ ($g\neq 0$) and $0 < \epsilon \le 0.1$. 
It is well-known that in the FHN model, the variables have no direct physiological interpretation. However, for the parameter ranges quoted above, the qualitative behaviour of the $x$'s and $y$'s are similar to that of the voltage and gating variables in the Hodgkin-Huxley equations.

\paragraph{Stationary points in the FHN system}

Neurons within the central nervous system can exist in a variety of dynamical states.  Many, for example, are in a state of quiescence and elicit a relaxation oscillation, known as an \emph{action potential}, when perturbed with a suprathreshold stimulus. This is oft termed as ``spiking''. With the appropriate choice of parameters, the FHN model can exhibit such behaviour.

However, upon varying these parameters, one can also invoke a \emph{Hopf bifurcation} resulting in relaxation oscillations at an intrinsic frequency without the need for any stimulation. Such self sustained neurons can be found within the central nervous system and can have complex interactions with the environment. A typical example is that of \emph{circadian cells} which act like the organisms clock cells and have been shown to have the ability to entrain their oscillations with the environments light-dark cycle \cite{veleri03}.

Neurons have also been shown to express bistability 
\cite{tasaki59},which again is present in the FHN model due to the cubic nonlinearity present in the system.

The dynamical states in the FHN model described above can be identified via a stability analysis of the stationary points in the system corresponding to $\frac{d x_{i}}{d t} = \frac{d y_{i}}{d t} = 0$. The solutions to this system shall henceforth be referred to as simply the steady states. Such an analysis is thus essential if one is to have a basic understanding of the system.  In fact, the steady state equations of this model have already been solved exactly for the $n=1$ and $n=2$ cases \cite{campbell01}. Recently a conjecture relating the steady states of a system and its Kolmogorov-Sinai entropy has brought to greater light the importance of steady state analysis for neural modeling in general\cite{baptista08,baptista09,baptista09b,baptista08b}.

It should be noted that computational and numerical algebraic geometry methods have found many applications in many branches of theoretical physics in general (see, e.g., Refs.~\cite{Mehta:2009zv, Mehta:2009, Gray:2008zs}). Below we use the DV method to solve the corresponding equations for the $n=3$ case and study the stability and bifurcation structure of these systems. The functions we use are in the packages \texttt{Groebner} and \texttt{RootFinding[Parametric]} of the computer algebra system \texttt{Maple 13}. Due to the polynomial-like nonlinearity also exhibited in the HR, ML and Izhikevich models, the same technique may also be applied to these systems.

The problem addressed in this paper can also be related to the more general problem of the algebraic analysis of the solutions of a differential system, studied for example in \cite{BLLMUab07}, \cite{NWmcs08},\cite{WXissac05} or \cite{YNAHcasc07},\cite{YNAHab07}.

It is also worth mentioning that, though we have performed our computations in \texttt{Maple}, some of the computations (quantifier elimination, sample points extraction) can be run using \texttt{Discoverer}\footnote{http://www.is.pku.edu.cn/~xbc/discoverer.html, developed by Wang, Xia et al.}, QEPCAD\footnote{http://www.usna.edu/Users/cs/qepcad/B/QEPCAD.html, developed by Hong et al.}, \texttt{REDUCE}\footnote{http://reduce-algebra.sourceforge.net/, developed by Hearn, Weispfenning et al.}, \texttt{Mathematica}\footnote{http://www.wolfram.com/products/mathematica/index.html}, \texttt{STRINGVACUA}\footnote{http://www-thphys.physics.ox.ac.uk/projects/Stringvacua/} or \texttt{RAG}\footnote{http://www-spiral.lip6.fr/~safey/RAGLib/distrib.html, developed by Safey El Din}.

\paragraph{Algebraic tools}



For this study, we used the Discriminant Variety~\cite{LRjsc07} and Cylindrical Algebraic Decomposition~\cite{Cbook75}. The Discriminant Variety is an implicit representation of the desired partition, while Cylindrical Algebraic Decomposition describes explicitly each cell of the partition.

Combined together, these two methods are well adapted to the analysis of the steady states (stable steady states). They provide a partition of the parameter space into connected cells, such that within each cell, the number of steady states (stable steady states) is constant.

Finally, the main algebraic criterion to decide if a solution is stable is the Routh-Hurwitz criterion, or its Li\'enard-Chipart variant \cite{LTbook85}. For our problem, we used a reduced criterion, more adapted to the algebraic DV and CAD methods.

\paragraph{Results}
The model in Eq.~\ref{eq:n_FHN_general_eq} has already been studied for the case $n=2$ in Ref.~\cite{campbell01}. We therefore focussed our work on the case $n\geq 3$.

\begin{figure}
\begin{minipage}{0.3\linewidth}
\begin{center}
\includegraphics[angle=0,width=\textwidth]{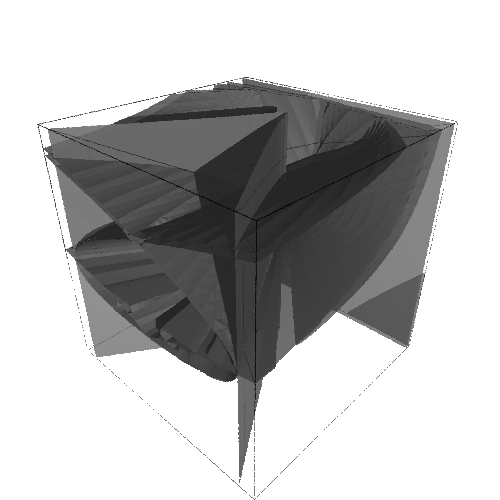}
 \end{center}
\caption{ Parameter space decomposition adapted to the number of steady states for $n=3$: in each connected component outside the surface, there is a constant number of steady states.}
\label{dv3d}
\end{minipage}
\hspace{0.05\linewidth}
\begin{minipage}{0.6\linewidth}
\tiny
\begin{tabular}{|c|@{}c@{}|@{}c@{}|@{}c@{}|@{}c@{}|@{}c@{}|@{}c@{}|}
     \hline g & $]-\infty,\mu_{1}[$ & $]\mu_{1},\mu_{2}[$ & $]\mu_{2},\mu_{3}[$ & $]\mu_{3},\mu_{4}[$ & $]\mu_{4},\mu_{5}[$ & $]\mu_{5},\mu_{6}[$ \\
    \hline Steady states & 3 & 3 & 3 & 3 & 3 & 3 \\
    \hline Stable states & 2 & 2 & 2 & 2 & 2 & 2 \\
    \hline
    \multicolumn{6}{c}{ }\\

     \hline g & $]\mu_{6},\mu_{7}[$ & $]\mu_{7},\mu_{8}[$ & $]\mu_{8},\mu_{9}[$ & $]\mu_{9},\mu_{10}[$ & $]\mu_{10},\mu_{11}[$ & $]\mu_{11},\mu_{12}[$ \\
    \hline Steady states & 3 & 3 & 3 & 3 & 3 & 15 \\
    \hline Stable states & 2 & 2 & 2 & 2 & 2 & 2 \\
    \hline
    \multicolumn{6}{c}{ }\\

     \hline g & $]\mu_{12},\mu_{13}[$ & $]\mu_{13},\mu_{14}[$ & $]\mu_{14},\mu_{15}[$ & $]\mu_{15},\mu_{16}[$ & $]\mu_{16},\mu_{17}[$ & $]\mu_{17},\mu_{18}[$ \\
    \hline Steady states & 15 & 15 & 15 & 15 & 27 & 27 \\
    \hline Stable states & 2 & 2 & 2 & 2 & 8 & 8 \\
    \hline
    \multicolumn{6}{c}{ }\\

     \hline g & $]\mu_{18},\mu_{19}[$ & $]\mu_{19},\mu_{20}[$ & $]\mu_{20},\mu_{21}[$ & $]\mu_{21},\mu_{22}[$ & $]\mu_{22},\mu_{23}[$ & $]\mu_{23},\mu_{24}[$ \\
    \hline Steady states & 27 & 27 & 27 & 27 & 27 & 27 \\
    \hline Stable states & 8 & 8 & 8 & 14 & 6 & 6 \\
    \hline
    \multicolumn{6}{c}{ }\\

    \cline{1-5} g & $]\mu_{24},\mu_{25}[$ & $]\mu_{25},\mu_{26}[$ & $]\mu_{26},\mu_{27}[$ & $]\mu_{27},\infty[$ \\
    \cline{1-5} Steady states & 27 & 27 & 15 & 15 \\
    \cline{1-5} Stable states & 6 & 6 & 6 & 6 \\
    \cline{1-5} \end{tabular}

     where:
\tiny

    $\mu_{1}\approx-1.690122506$, $\mu_{2}\approx-.3010608116$, $\mu_{3}\approx-.2787081900$, $\mu_{4}\approx-.2530108085$, \\
    $\mu_{5}\approx-.2200916671$, $\mu_{6}\approx-.2119432516$, $\mu_{7}\approx-.1972651791$, $\mu_{8}\approx-.1944444444$, \\
    $\mu_{9}\approx-.1797585308$, $\mu_{10}\approx-.1768496368$, $\mu_{11}\approx-.1666666667$, $\mu_{12}\approx-.1111111111$, \\
    $\mu_{13}\approx-.09533095509$, $\mu_{14}\approx-.08765527212$, $\mu_{15}\approx-.08093319477$, $\mu_{16}\approx-.06753407911$, \\
    $\mu_{17}\approx-.06313592910$, $\mu_{18}\approx.1646239760$, $\mu_{19}\approx.2210124208$, $\mu_{20}\approx.2847525372$, \\
    $\mu_{21}\approx.3136620682$, $\mu_{22}\approx.3333333333$, $\mu_{23}\approx.3372096920$, $\mu_{24}\approx.3373316121$, \\
    $\mu_{25}\approx.3483417636$, $\mu_{26}\approx.3562092137$, $\mu_{27}\approx.3888888889$.

\caption{Number of steady and stable states for $a=0$, $b=2$ according to the parameter $g$.}
\label{stable3}
\end{minipage}

\end{figure}

First, we succeeded in describing the steady steady states for $n=3$. The result of our computation is summarized in Figure \ref{dv3d}.

However, when $n\geq 3$, the computations of the stable states are much more difficult and we did not succeed in describing the full parameter space according to the number of stable steady states. Nevertheless, by fixing the parameters $a$ and $b$ to the numerical values (respectively $0$ and $2$), we were able to describe the number of steady and stable steady states according to the free parameter $g$, for the $n=3$ case. Since there is only one parameter, the description of the parameter space is a union of intervals in the parameter $g$ for which the number of stable (and steady) solutions is constant.

The results of the computations are summarized in Figure \ref{stable3}.

\paragraph{Acknowlegement}
DM was supported by Science Foundation of Ireland. SP was in part supported by the Irish Research Council for Science Engineering and Technology (IRCSET).


\footnotesize

\bibliography{biblioabstract}

\end{document}